\documentclass[pre,amsmath,amssymb,twocolumn,nofootinbib,floatfix]{revtex4-2}
\usepackage[colorlinks=true,linkcolor=blue,urlcolor=blue,citecolor=blue]{hyperref}
\usepackage{graphicx}
\usepackage{bm}
\usepackage{times}
\usepackage{xcolor}

\newcommand{\kBT}{k_\mathrm{B}T}

\newcommand{\refcite}[1]{\mbox{Ref.\hspace{0.25em}\cite{#1}}}
\newcommand{\figref}[1]{\mbox{Fig.\hspace{0.25em}\ref{#1}}}
\newcommand{\Eqref}[1]{\mbox{Eq.\hspace{0.25em}\eqref{#1}}}
\newcommand{\Eqsref}[1]{\mbox{Eqs.\hspace{0.25em}\eqref{#1}}}
\newcommand{\Appref}[1]{\mbox{Appendix\hspace{0.25em}\ref{#1}}}
\newcommand{\chiC}{\chi_\mathrm{c}}

\newcommand{\rmd}{\mathrm{d}}

\newcommand{\vect}{\boldsymbol}

\graphicspath{{figures/},{}}

\begin{document}

\title{Intermediate physical interactions induce spatiotemporal dynamics in Turing patterns}

\author{Cathelijne ter Burg}
\affiliation{Max Planck Institute for Dynamics and Self-Organization, Am Faßberg 17, 37077 Göttingen, Germany}

\author{David Zwicker}
\affiliation{Max Planck Institute for Dynamics and Self-Organization, Am Faßberg 17, 37077 Göttingen, Germany}

\begin{abstract}

Turing patterns are a central paradigm for describing spatial patterns in nature. The corresponding theory of reaction-diffusion dynamics combines ideal diffusion with nonlinear reactions, resulting in patterns when species diffuse at different rates and reactions are sufficiently nonlinear. However, real systems are more complex and particularly involve physical interactions between constituents. While such interactions can promote patterns, we here show that they can also induce dynamic, chaotic patterns. These patterns exhibit well-defined length and time scales, which result from cycles of droplet coarsening and fission. The dynamical patterns combine properties of traditional Turing patterns and chemically active droplets, which emerge for strong physical interactions. Our analysis thus reveals three qualitatively different regimes that emerge when two components interact physically and undergo nonlinear reactions.

\end{abstract} 

\maketitle


\section{Introduction}


Reaction-diffusion systems have been a cornerstone of theoretical modeling of natural patterns since Alan Turing introduced them to explain patterns of morphogens~\cite{Turing1952}.
These \emph{Turing patterns} form when a locally self-enhancing activator is counterbalanced by a faster-diffusing inhibitor that suppresses production globally~\cite{Vittadello2021,Kondo2010}.
Linear stability analysis reveals that patterns emerge only if (i) the inhibitor diffuses much faster than the activator, and (ii) the underlying reaction kinetics are strongly nonlinear.
If these conditions are met, the framework explains various natural patterns ranging from nano-crystals~\cite{Fuseya2021}, complex tissues~\cite{RechoHallouHannezo}, to geophysical phenomena~\cite{Goehring2004}.

Reaction-diffusion systems provide a mathematically elegant description of pattern formation, but the underlying assumptions are often questionable in real systems.
In particular, to exhibit the necessary nonlinearities in the reactions, the involved components typically interact physically, but these interactions would also affect spatial transport~\cite{Menou2023}.
Strong physical interactions can even give rise to phase separation, which has been shown to be relevant in the spatial organization of biological cells~\cite{Banani2017,Tsai2022,Liu2013,Choi2020,Dignon2020,ZwickerPaulinTerBurg2025}.
The resulting droplets can then be influenced by reactions, resulting in regular pattern reminiscent of Turing patterns~\cite{Glotzer1994,Christensen1996,ZwickerOstwald2015,Carati1997,Zwicker2022a,ZwickerPaulinTerBurg2025}.
However, even weak interactions can broaden the parameter regime where Turing patterns emerge~\cite{Menou2023} and lower the energetic costs required to maintain them~\cite{terBurgZwicker2025}.
These earlier investigations of the role of physical interactions onto pattern formation focused on regular, stationary patterns, and it is currently unclear whether the interplay of nonlinear reactions and physical interactions might also lead to dynamic states.


We here present a systematic parameter study of a reaction-diffusion system of two components with physical interactions.
Our simulations reveal three qualitatively different patterns, which resemble classical Turing patterns at weak interactions (TP regime), chemically active droplets at strong interactions (AD regime), and an intermediate dynamic regime exhibiting spatiotemporal dynamics (DP regime).
Using stability analysis and numerical simulations, we rationalize how parameters affect the emerging length scale and we show that the dynamical regime emerges from a perpetual cycle of coarsening and fission of droplets.
 
 \section{Model}

To study Turing patterns with physical interactions, we consider an isothermal mixture of an activator~$A$ and an inhibitor~$I$, which are embedded in a solvent $S$.
We consider an incompressible system, whose composition is described by the respective volume fractions~$\phi_A$ and $\phi_I$, while the solvent fraction is given by  $\phi_S = 1-\phi_A-\phi_I$. 
The dynamics of the system are governed by two coupled partial differential equations for the concentration fields~\cite{Menou2023},
\begin{align}
	\partial_t \phi_i &= \nabla \cdot  (D_i \phi_i \nabla \bar{\mu}_i) +  s_i
		\;, & i &= A, I
	\label{eqn:pde}
\end{align}
where the first term on the right accounts for diffusive fluxes proportional to  diffusivities~$D_i$, and the source term~$s_i$ captures reactions.
The diffusive fluxes scale with the fraction~$\phi_i$, and we neglect explicit cross-diffusion, essentially assuming fast solvent dynamics~\cite{Kramer1984,ZwickerPaulinTerBurg2025}.
The diffusive fluxes are driven by gradients in non-dimensional exchange chemical potentials~$\bar{\mu}_i$ derived from the free energy $F$ of the system, $\bar{\mu}_i = (\nu/\kBT) \delta F/\delta \phi_i$, where $\kBT$ is the relevant energy scale and $\nu$ denotes the molecular volume, which we take to be the same for all components.
For simplicity, we choose the free energy~\cite{Safran2018,Rubinstein2003,Cahn1958}
\begin{align}
	F[\phi_A, \phi_I]  &= \frac{\kBT}{\nu} \int\Bigl[
		\phi_A \ln\phi_A + \phi_I \ln\phi_I + \phi_S \ln\phi_S  
\notag\\
		 & +    \chi \phi_A \phi_I  + \frac{w^2}{2} \left( |\nabla \phi_A|^2 + |\nabla \phi_I|^2\right)
	\Bigr] \rmd V
	\;, 
	 \label{eqn:free_energy}
\end{align}
which originates from Flory-Huggins theory~\cite{Flory1942,Huggins1941}.
It accounts for translational entropies of all components (first line) and interactions that can lead to phase separation (second line).
In particular, the term proportional to the Flory parameter~$\chi$ accounts for physical interactions between activator and inhibitor, and the last term penalizes interfaces, which will then exhibit a typical width~$w$.
The free energy~$F$ leads to segregation of $A$ from $I$ for positive $\chi$, whereas negative $\chi$ leads to attraction.
Absent interactions ($\chi=0$) corresponds to ideal diffusion commonly used for Turing patterns~\cite{Menou2023}, where spatial fluxes tend to homogenize the system.
To form patterns in this typical case of ideal diffusion, the chemical reactions described by $s_i$ in \Eqref{eqn:pde} need to be sufficiently nonlinear.
To describe such nonlinearities generically, we consider 
\begin{align}
	s_i = k \cdot \left[ \frac{2 \phi_0}{1+(\phi_I / \phi_A)^h} - \phi_i \right]  
	\;,
	\label{eqn:reaction_rate} 
\end{align}
which provides separate control over the reaction rate $k$, the fraction~$\phi_0$ in the homogeneous steady state, and the reaction nonlinearity~$h$.
This reaction scheme corresponds to the strongly driven limit of the system we studied in \refcite{terBurgZwicker2025}, and was earlier shown to  produce patterns in the absence of interactions ($\chi=0$)~\cite{Menou2023}.

\begin{figure*} 
	\includegraphics[width=\linewidth]{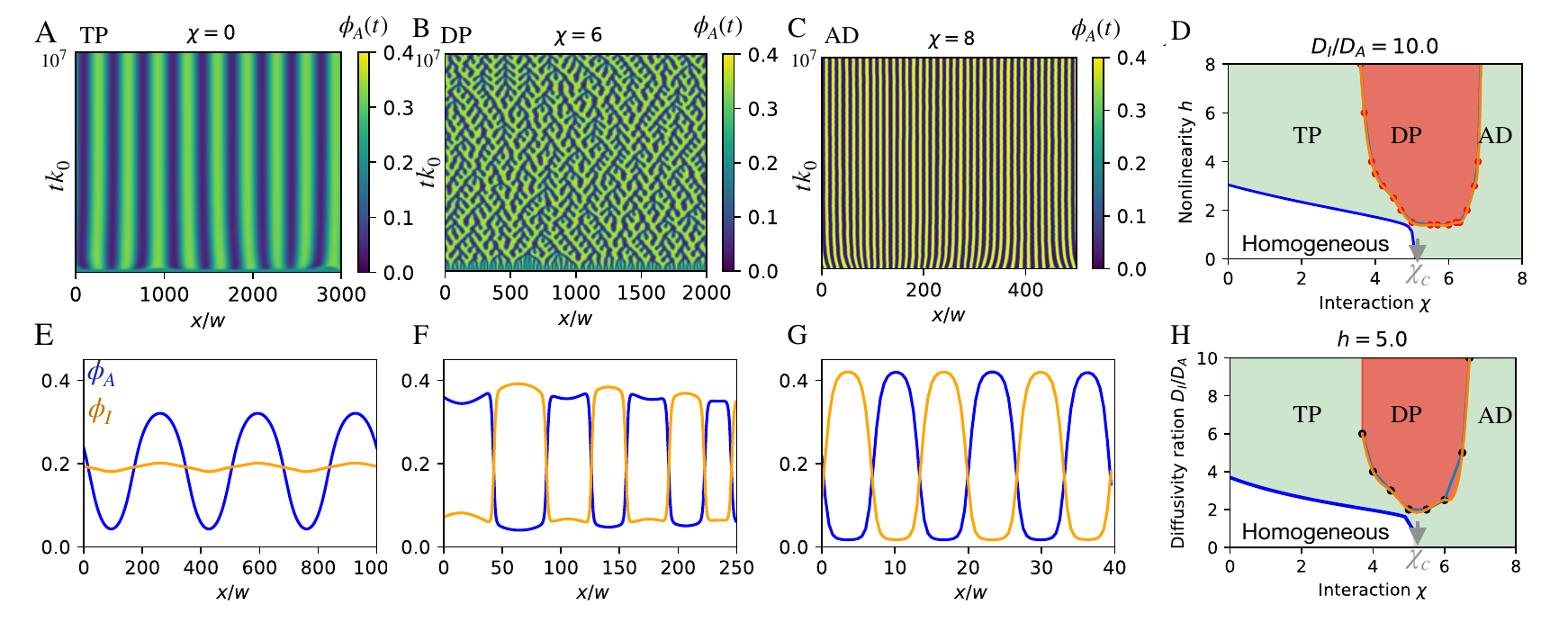}  
	\caption{\textbf{A dynamical regime (DP) separates stationary Turing patterns (TP) from active droplets (AD).} (A--C) Activator volume fraction $\phi_A$ as a function of the one spatial coordinate~$x$ and time~$t$ for the Turing pattern regime (TP, $\chi=0$), the dynamical pattern regime (DP, $\chi=6$), and the active droplet regime (AD, $\chi=8$). 
	(E--G) Activator fraction $\phi_A$ (blue) and inhibitor fraction $\phi_I$ (orange) as a function of $x$ at $t=10^7 \, k_0^{-1}$ corresponding to panels A--C.
	(D, H) State diagrams indicating the three regimes as a function of interaction strength $\chi$ and reaction nonlinearity $h$ (panel D) and diffusivity ratio $D_I/D_A$ (panel H).
	Homogenous states are stable in the white region.
	(A--H) Model parameters are $h = 5$, $D_I/D_A = 10$, $k/k_0 = 10^{-3}$, 
$k_0 = D_A/w^{2}$, $\phi_0 = 0.2$, and system size $L=2000\,w$; Numerical details are described in \Appref{app:numerics}.
}
	\label{fig:kymographs}
\end{figure*}

We have previously shown that the system described by \Eqsref{eqn:pde}--\eqref{eqn:reaction_rate} exhibits patterns if the reaction nonlinearity~$h$ and the diffusivity ratio $D_I/D_A$ are sufficiently large~\cite{Menou2023}.
Moreover, the minimal values  of these quantities required for pattern formation are lower for larger repulsion~$\chi$ between activator $A$ and inhibitor~$I$.
However, the emerging length scale~$\ell$ of the resulting patterns remained difficult to interpret and seemed to exhibit a transition in qualitative behavior when $\chi$ crossed the value required for phase separation without reactions~\cite{Menou2023}.


\section{Results}

To analyze the influence of the interaction parameter~$\chi$ on the resulting patterns, we performed  numerical simulations in one-dimensional systems.
We visualize the dynamics in space-time plots, which confirm that regular patterns form for weak interactions ($\chi=0$, \figref{fig:kymographs}A) and strong interactions ($\chi=8$, \figref{fig:kymographs}C).
However, we also discovered an additional dynamical regime for intermediate interactions ($\chi=6$, \figref{fig:kymographs}B), which is not characterised by steady state patterns.
The fact that this dynamical regime occurs for intermediate interaction strengths suggests that the two stationary patterns might also be qualitatively different.
Indeed, for low interactions, only the activator $A$ exhibits strong variations, whereas the inhibitor $I$ is distributed almost homogeneously (\figref{fig:kymographs}E).
This behavior is consistent with the paradigm of \emph{local activation and global inhibition} that explains Turing patterns, suggesting these patterns are in the classical Turing pattern regime (TP)~\cite{Turing1952, Gierer1972}.
In contrast, strong interactions lead to segregation of $A$ and $I$ (\figref{fig:kymographs}G), which is reminiscent of phase separation.
This regime corresponds to active droplets (AD), where patterns form by phase separation, but their coarsening is suppressed by chemical reactions~\cite{Glotzer1994,Christensen1996,ZwickerOstwald2015,Kirschbaum2021,ZwickerPaulinTerBurg2025}.
These two stationary regimes are separated by the dynamical regime (DR) at intermediate interaction strengths~$\chi$.

The three qualitatively different patterns are clearly revealed when additional parameters, like the nonlinearity~$h$ or the diffusivity contrast $D_I/D_A$ are varied.
\figref{fig:kymographs}D and \figref{fig:kymographs}H show that the dynamical regime only emerges for intermediate $\chi$ in systems where pattern formation is sufficiently strong.
The relevant interaction strength lies in the vicinity of the critical interaction strength $\chiC$, which is required to induce phase separation when reactions are absent.
Linear stability predicts  $\chiC=\phi_0^{-1}$~\cite{Menou2023}, which implies $\chiC=5$ for the data in \figref{fig:kymographs}.
This already suggest that the dynamical regime emerges from an interplay of weak phase separation and strong reactions.
To understand the different regimes in detail, we next investigate the extreme cases of weak and strong interactions, before we study the dynamical regime.



\subsection{Weak  physical interactions lead to Turing patterns (TP)}
\label{sec:turing_pattern}

Weak interactions ($\chi\ll\chiC$) cannot induce phase separation in systems without reactions.
Consequently, we expect that the system with reactions effectively behaves as a reaction–diffusion system with ideal diffusion.
In this case, interactions captured by $\chi$ lead to ideal cross-diffusion~\cite{Menou2023,terBurgZwicker2025}, which enhances pattern formation and thus broadens the parameter range where stationary patterns emerge.
In this weak-interaction regime, the activator forms a periodic steady-state profile of a well-defined wavelength~$\ell$, whereas the inhibitor profile remains nearly flat, exhibiting only small variations (\figref{fig:kymographs}E).
In particular, these patterns can be studied by linear stability analysis, which predicts stationary patterns above threshold values for $h$ and $D_I/D_A$~\cite{Menou2023}, consistent with our numerical data (\figref{fig:kymographs}D, H).
Moreover, the most unstable mode provides a prediction for the pattern length scale~$\ell$.
In the limit of a large diffusivity contrast ($D_I \gg D_A$), we find $\ell_\mathrm{TP} \approx 4\pi \sqrt{D_A/(kh)}$~\cite{Menou2023}, suggesting that the length scale $\ell_\mathrm{TP}$ in the Turing regime is governed by the reaction-diffusion length $\sqrt{D_A/k}$ associated with the activator.
Taken together, the limit of weak interactions is consistent with traditional reaction-diffusion systems, where interactions merely introduce some cross-diffusivity.


\subsection{Strong physical interactions lead to active droplets (AD)}
\label{sec:AD}

In the limit of strong interactions ($\chi\gg\chiC$), diffusive fluxes dominate the behavior and enforce phase separation, which is characterized by coexistence of activator-rich droplets with inhibitor-rich surrounding regions (\figref{fig:kymographs}G).
Without reactions, the dynamics of such a system would be determined by the gradient term proportional to $w^2$ in \Eqref{eqn:free_energy}, which describes interface-penalizing surface tension.
Consequently, such a system exhibits a coarsening process known as Ostwald ripening~\cite{Voorhees1985,Voorhees1992}, which in one dimensions implies that the pattern length scale $\ell$ grows logarithmically in time.
Chemical reactions can suppress this coarsening by providing fluxes that counteract the effect of surface tension~\cite{ZwickerOstwald2015}.
In the simplest case of such chemically active droplets, $\ell$ scales with the interfacial width $w$ and a reaction-diffusion length~$l$.
In particular, binary models with just one diffusivity result in $\ell\sim w^{1/3}l^{2/3}$, which implies $\ell\sim k^{-1/3}$~\cite{Glotzer1994,Christensen1996,ZwickerOstwald2015,ZwickerPaulinTerBurg2025}.
In our more complex model, we have two distinct reaction-diffusion lengths, associated with the activator, $l_A=\sqrt{D_A/k}$, and the inhibitor, $l_I =\sqrt{D_I/k}$, respectively.
Numerical evidence shown in Appendix~\ref{app:length_scaling} suggests that the length scale $\ell_{\rm AD}$ in the active droplet regime scales as $\ell_{\rm AD} \sim (wl_Al_I)^{1/3}$, so that both reaction-diffusion lengths contribute equally.
Moreover, this form maintains the scaling $\ell \sim k^{-1/3}$ observed in simpler systems~\cite{Glotzer1994,Christensen1996,ZwickerOstwald2015}.
These observations suggest that the strong interactions in the active droplet regime lead to chemically active droplets, which are qualitatively different from the Turing patterns at weak interactions.
Between these two regimes, we found the dynamical regime, which might inherit properties from both adjacent regimes.

\subsection{Intermediate physical interactions lead to dynamical patterns (DP) involving cycles of droplet growth and fission}
\label{sec_DR}

\begin{figure*} 
	\includegraphics[width=\linewidth]{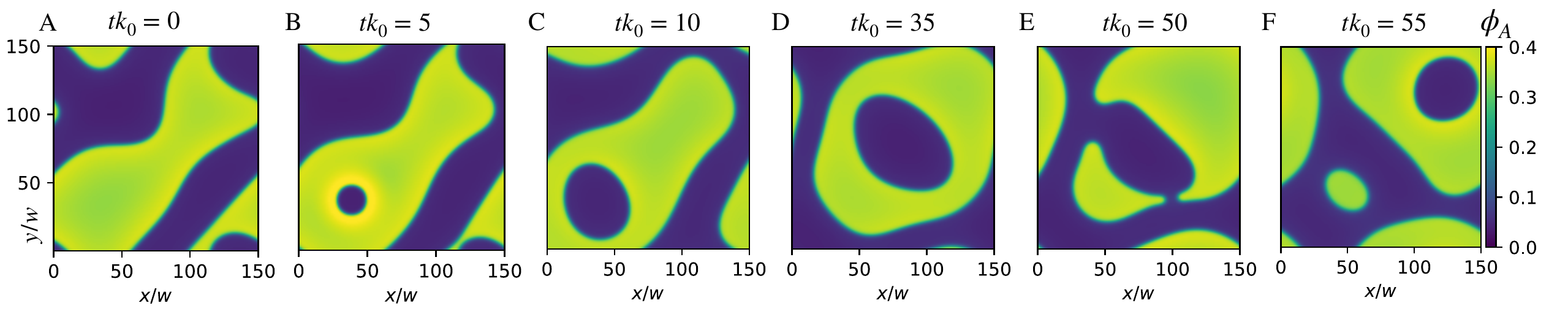}  
	\caption{\textbf{Dynamical regime exhibits cycles of growth and fission.}
	Snapshots of the activator fraction $\phi_A$ of a two-dimensional system for the indicated times $t$.
	Model parameters are $\chi = 6$, $h = 5$, $D_I/D_A = 3$, $k = 0.01\,k_0$, $k_0 = D_A/w^2$, and $\phi_0= 0.2$.	}
	\label{fig:2d_simulations}
\end{figure*}

For intermediate interactions ($\chi\sim\chiC$), our numerical simulations suggest that the system does not reach a stationary pattern and instead develops spatiotemporal dynamics with a dominant length- and time-scale (\figref{fig:kymographs}B).
The nature of the dynamics becomes even more clear in two-dimensional simulations:
\figref{fig:2d_simulations} shows snapshots of the key events, where activator-rich droplets (yellow region) develop an activator-poor bubble (small blue region), which then grows and induces fission of the activator-rich droplet.
Evidently, these cycles of growth and fission give rise to the dynamic patterns.
Here, droplet growth could be a natural consequence of Ostwald ripening~\cite{Voorhees1985,Voorhees1992} if reactions are not strong enough to suppress it.
Alternatively, droplet growth might also be driven by chemical reactions~\cite{ZwickerPaulinTerBurg2025}.
In contrast, the spontaneous fission is qualitatively different from previously observed droplet division~\cite{Zwicker2017}.
Instead, fission is preceded by the formation of a bubble, which is reminiscent of vacuole formation observed in chemically active droplets~\cite{Bergmann2023,Verstraeten2025}, although spontaneous fission was not observed in these systems.



\begin{figure*} 
	\includegraphics[width=\linewidth]{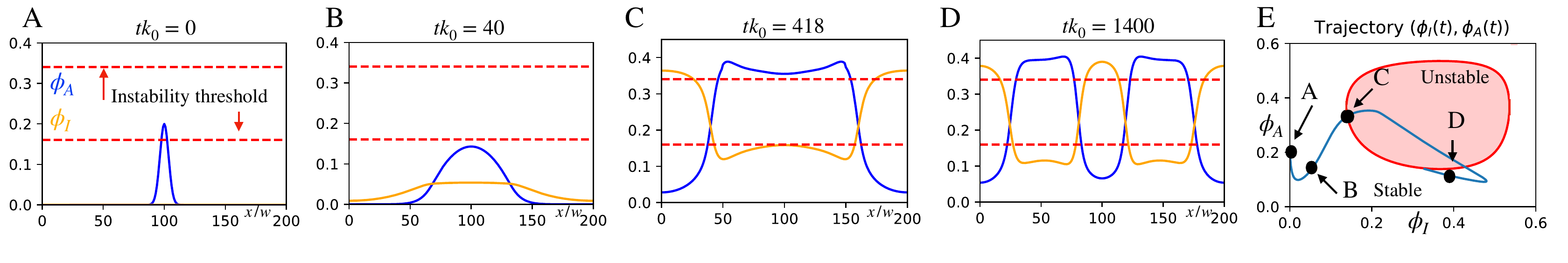}  
	\caption{\textbf{Growth of a single droplet reveals fission mechanism.}
	(A--D) Volume fractions of activator ($\phi_A$, blue lines) and inhibitor ($\phi_I$, orange lines) at four indicated times $t$.
	Red dashed lines mark the instability boundary that is crossed in panel C.
	(E) Phase diagram of system without reactions ($k=0$) as a function of $\phi_A$ and $\phi_I$.
	Homogeneous states are unstable in the red region (Appendix~\ref{app:stability_analysis}).
	Blue line shows trajectory of $\phi_A(t)$ and $\phi_I(t)$ at the droplet center at $x = 100\,w$ with points marking states shown in panels A--D. 
	(A--E) Model parameters are $\chi = 4.5$, $k = 0.01\,k_0$, $D_I/D_A = 10$, $h = 5$, $\phi_0 = 0.2$, and $k_0 = D_A/w^2$.
	}
	\label{fig:isolated_droplet}
\end{figure*}

To understand fission in detail, we consider the dynamics of a single, isolated droplet in one dimension by initializing a system with a narrow peak of activator~$A$ (\figref{fig:isolated_droplet}A).
The dynamics given by \Eqref{eqn:pde} then lead to production of $A$ and $I$ (\figref{fig:isolated_droplet}B), implying that the droplet grows.
Because the inhibitor diffuses faster than the activator ($D_I \gg D_A$), the inhibitor profiles are broader (orange lines) while the activator profiles exhibit larger peaks (blue lines).
Growth continues until the inhibitor fraction reaches the instability line (red dashed line in \figref{fig:isolated_droplet}C), which indicates the composition at which a homogeneous state becomes linearly unstable (Appendix~\ref{app:stability_analysis}).
The instability induces rapid dynamics, where particularly the activator profile dips (\figref{fig:isolated_droplet}D), while the inhibitor rises, thus explaining how the bubble forms.
Consequently, fission completes and the two resulting activator droplets can start a new cycle of growth and fission.


The fission dynamics can also be visualized in the general phase space of the two components (\figref{fig:isolated_droplet}E).
A linear stability analysis shows that homogeneous states are unstable in the red region (Appendix~\ref{app:stability_analysis}).
Tracing the droplet mid-point (blue line) then reveals that the droplet initially remains in the stable region until the particular time points corresponding to \figref{fig:isolated_droplet}C.
Further growth then triggers the instability, which is visible by the short excursion into the unstable region.
Once splitting took place (\figref{fig:isolated_droplet}D), the mid point leaves the unstable region, evolving toward a new configuration.
Further growth will then trigger another instability, driving a perpetual cycle of growth and fission.


An important feature of the dynamics is that right prior to hitting the instability, the activator profile exhibits a dip in the center of the droplet (\figref{fig:isolated_droplet}C).
This is surprising since both the activator and the inhibitor are produced in the activator-rich droplet, so one could naively expect that they both exhibit a peak inside the droplet.
However, the actual profiles are also affected by diffusive fluxes, and particularly cross-diffusion, which is induced by physical interactions~\cite{Menou2023,terBurgZwicker2025}.
For intermediate interactions relevant to the DP regime, we can approximate the diffusive flux by expanding the term $\nabla( D_i \phi_i \nabla \bar\mu_i)$ in \Eqref{eqn:pde} in the volume fractions.
For the diffusive flux of the activator, this results in a term $\chi D_A  \nabla(\phi_A \nabla \phi_I)$, indicating that its flux is affected by the fraction of the inhibitor, which is cross-diffusivity.
This cross-diffusion effectively pushes the activator out of its dense phase, leading to the observed depletion in the center (\figref{fig:isolated_droplet}C).
Furthermore, the central depletion of $A$ induces a cross-diffusion flux for the inhibitor $I$, which thus accumulates in the center, amplifying the total effect.
This picture is consistent with the observation that the dynamic pattern only emerges when the diffusivity of $I$ is sufficiently large compared to $A$ (\figref{fig:kymographs}H).
Taken together, we  conclude that cross-diffusivity induced by the interactions triggers the instability in isolated droplets, and we hypothesize that this effect also explains droplet splitting when there are many droplets.
Indeed, we show in Appendix \ref{app:instability_full_model} that the qualitative features of this analysis for an isolated droplet are reproduced in the full dynamics containing many droplets.

We identified a scenario where droplet growth leads to an instability, which induces fission.
The resulting smaller droplets can then grow again, and the cycle continues.
In a dense system of many droplets, droplet growth might be hindered by surrounding droplets, so that droplet growth must happen at the expense of shrinkage of other droplets, which is similar to the coarsening process of Ostwald ripening~\cite{Voorhees1985,Voorhees1992}.
In the long term limit depicted in \figref{fig:kymographs}B, we thus see competition of neighboring droplets, and only the large droplets split spontaneously (Supplemental \figref{fig:instability_full_model}).
Small disturbances affect which droplets win the competition, which explains the tree-like patterns observed in \figref{fig:kymographs}B.
Note that the onset of the dynamical regime can be quite delayed (\Appref{app:transients}), suggesting that a regular periodic state (similar to what we observed in the active droplet regime) can be long-lived.

\subsection{Interactions and reaction rates control pattern length scale}
\label{sec_Length_Scale}

\begin{figure*}
    \includegraphics[width=\linewidth]{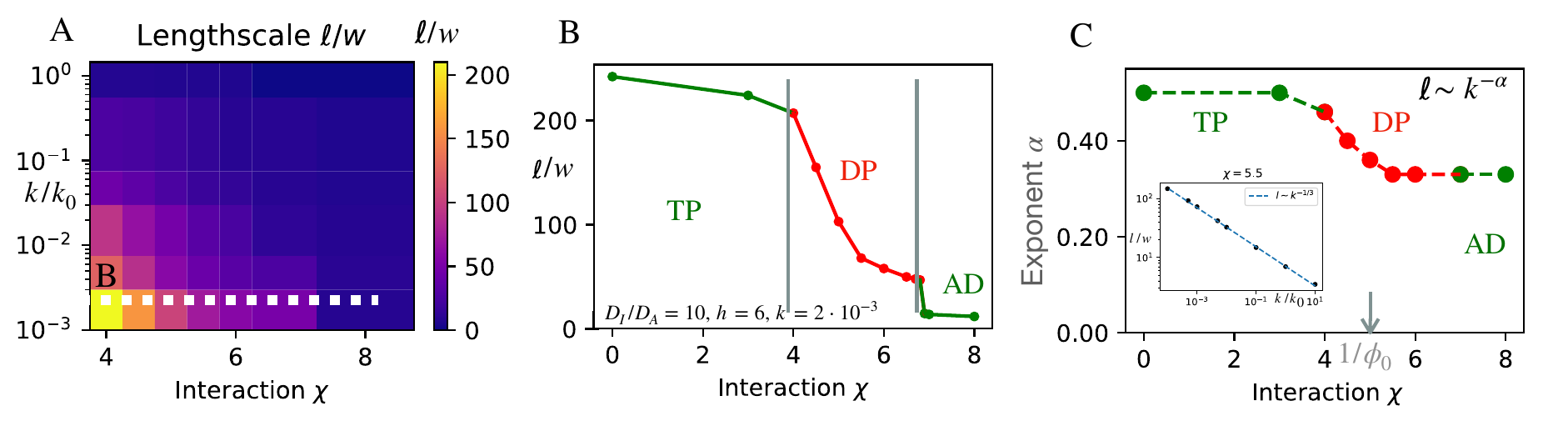}
    \caption{
	\textbf{Interactions and reaction rates control pattern length scale.}
	(A) Pattern length scale~$\ell$ (measured from time-averaged structure factor; see Appendix~\ref{app:length_scale}) as a function of interaction strength $\chi$ and reaction rate $k$.
	White dotted line corresponds to panel B.
	(B) $\ell$ as a function of $\chi$ for $k = 2 \cdot 10^{-3}\, k_0$. 
	(C) Scaling exponent $\alpha$ extracted by fitting $\ell \sim k^{-\alpha}$ on a log-log scale to data in panel A as a function of $\chi$ (inset shows data for $\chi=5.5$).
	(A--C) Model parameters are $D_I/D_A = 10$, $h = 6$, $\phi_0 = 0.2$, $k_0 = D_A/w^2$, and $tk_0 = 10^7$.
	}
    \label{fig:length_scale}
\end{figure*}

We showed that the dynamical pattern regime (DP) lies between the Turing pattern regime (TP) and the active droplet regime (AD).
The two stationary regimes (TP and AD) exhibit very different dependencies of the observed pattern wave length $\ell$ on the parameters, most notably the reaction rate $k$. 
We showed above that $\ell_\mathrm{TP} \sim k^{-1/2}$, whereas $\ell_\mathrm{AD} \sim k^{-1/3}$, which raises the question of how $\ell$ behaves in the intermediate DP regime.
To answer this question, we performed numerical simulations for different values of $k$ and measured the pattern length scale $\ell$ from the spatial mean of the structure factor (Appendix~\ref{app:length_scale}).
\figref{fig:length_scale}A shows that $\ell$ generally decreases with larger $k$ and larger interaction strength~$\chi$.

To understand how the pattern length scale~$\ell$ changes with $\chi$, and particularly between the three regimes we identified, we next consider the fixed rate $k=2\cdot10^{-3}\,k_0$.
\figref{fig:length_scale}B shows that $\ell$ slightly decreases with increasing $\chi$ in the TP regime.
This decrease becomes much stronger in the DP regime, presumably reflecting the stronger influence of the interactions, which now drive phase separation.
The length scale~$\ell$ decreases further when we transition to the AD regime, which is accompanied by a jump in $\ell$, suggesting a  first-order transition.
This observation would further suggest that multistability is possible in the vicinity of the transition, which might explain the long transients that we observed in the DP regime (\figref{fig:transients}).

To investigate the influence of the reaction rate~$k$ on the pattern length scale~$\ell$, we next analyze the full data presented in \figref{fig:length_scale}A. 
In particular, we assume that the $k$-dependence can be captured by a power law, $\ell \sim k^{-\alpha}$, and determine the exponent $\alpha$ by fitting to the data on a log-log scale (inset of \figref{fig:length_scale}C).
\figref{fig:length_scale}C shows that $\alpha$ assumes the expected values in the stationary TP  ($\alpha\approx\frac12$) and AD  ($\alpha \approx \frac13$) regimes. The exponent smoothly interpolates between these values in the intermediate DP regime, which suggest that the dynamical regime indeed mixes aspects from Turing patterns and active droplets.
Taken together, this analysis demonstrates that reactions largely control the pattern length scale, but the details depend on the respective dynamical regime.

\section{Discussion}

We showed that physical interactions fundamentally alter pattern formation in reaction-diffusion systems.
Beyond stationary Turing patterns at weak interactions and active droplets at strong interactions, we identified an intermediate regime with intrinsically dynamical patterns.
In this regime, domains undergo repeated cycles of growth, instability, and fission, leading to persistent spatiotemporal dynamics rather than convergence to a steady state.
The observed growth--fission cycles resemble spatiotemporal patterns reported in other non-equilibrium systems, including chemotaxis models where transport processes destabilize large domains and prevent steady-state patterning~\cite{PainterHillen2011,Das2004,Das2005,Chakraborty2010,Frey2025}. 
Although the microscopic mechanisms differ, these systems share the feature that  instabilities limit coarsening and promote dynamic reorganization. 

The three regimes we identified exhibit qualitatively different behaviors of the pattern length scale~$\ell$.
In the Turing regime, we found $\ell \sim k^{-1/2}$, which is consistent with the idea that $\ell$ is controlled by a reaction-diffusion length scale.
We also recover the scaling $\ell \sim k^{-1/3}$ known for chemically active droplets~\cite{Glotzer1994,Christensen1996,ZwickerOstwald2015}, although we do not observe the scaling $\ell \sim k^{-1/4}$ predicted for weak phase separation~\cite{Christensen1996}.
Interestingly, we observe intermediate scaling relations in the dynamical patterns, suggesting a gradual shift in the dominant physical mechanism from a Turing instability to chemically active droplets as the physical interactions~$\chi$ are increased.

Our results demonstrate that components that interact physically and undergo nonlinear reactions can exhibit a wealth of dynamical behaviors.
The qualitative results are likely generic and do not depend on the particular choice of the diffusive fluxes or reaction kinetics, although this needs to be tested thoroughly.
In contrast, more complex behaviors are expected when additional complexity present in natural systems is added.
In particular, natural systems typically comprise many more components~\cite{Luo2023}, systems are spatially structured~\cite{Frey2025}, components might exhibit higher-order interactions~\cite{Luo2024}, and stochastic fluctuations matter~\cite{Ziethen2023}, which could all influence the observed behavior.
Indeed, in similar systems droplets have already been show to exhibit spontaneous splitting \cite{Zwicker2017}, the formation of liquid shells~\cite{Bergmann2023,Verstraeten2025}, and self-propulsion \cite{Demarchi2023,Qiang2025a}.
It will thus be paramount to study such systems in detail to understand the non-equilibrium spatio-temporal organization of soft and biological matter.

\textit{Acknowledgments---}We gratefully acknowledge funding from the Max Planck Society and the European Research Council (ERC) for financial support under the European Unions Horizon 2020 research and innovation programme (‘EmulSim’ with Grant Agreement No. 101044662).


\appendix 


\renewcommand\thefigure{S\arabic{figure}}
\setcounter{figure}{0}

\renewcommand{\theequation}{\thesection.\arabic{equation}}
\counterwithin*{equation}{section}
\setcounter{equation}{0}

\section{Numerical methods}
\label{app:numerics}
\label{app:length_scale}
We perform numerical simulations of \Eqref{eqn:pde} in the main text on a one-dimensional, equidistantly discretized grid with periodic boundary conditions using second-order finite differences to approximate differential operators~\cite{Zwicker2020python}.
We evaluate $\nabla \mu_i$ on a staggered grid to ensure material conservation.
We used an explicit Euler scheme for the time evolution with adaptive timestepping.
We typically use a spatial discretisation of two grid points per unit length $w$ and increased it to four when necessary.



\begin{figure}
	\includegraphics[width=\linewidth]{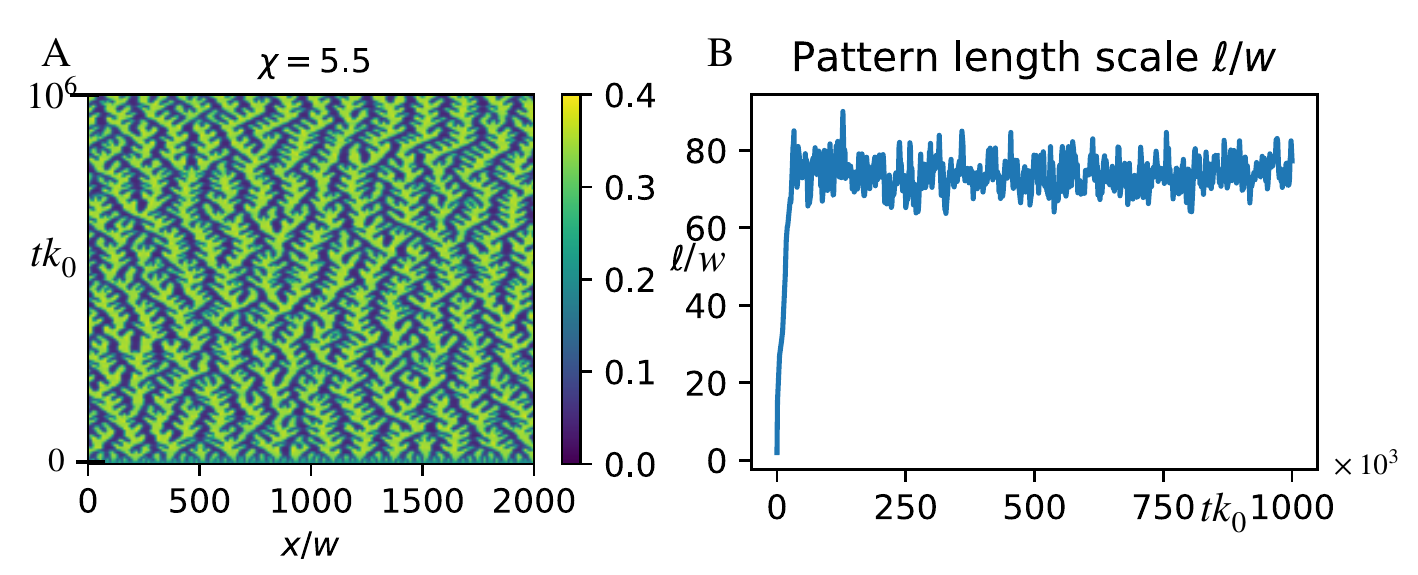}
	\caption{\textbf{Length scale determination in dynamic pattern regime requires time averaging.}
	(A)
	Activator volume fraction $\phi_A$ as a function of the single spatial coordinate $x$ and time $t$.
	(B)
	Pattern length scale $\ell$ (defined as mean structure factor) of data in panel A as a function of $t$.
	Model parameters are $k = 10^{-3}\,k_0$, $\chi = 5.5$, $D_I/D_A = 10$, $h = 5$, $\phi_0 = 0.2$, and $k_0 = D_A/w^2$.   }
	\label{fig_supp_Time_Av_Sq}
\end{figure}

Length scales reported in the main text are measured using the mean of the structure factor.
To compute this quantity, we determine the structure factor
\begin{equation}
	S(q,t) = \left|\hat{\phi}_A(q,t)\right|^2
	\;,
\end{equation}
at each time point, where $\hat{\phi}_A(q,t)$ denotes the spatial Fourier transform of the activator  fraction~$\phi_A(x, t)$. 
The average of the structure factor defines an instantaneous length scale $\ell(t)$,
\begin{align}
	\ell(t) &= \frac{2\pi}{\langle q \rangle_t}\;,
	&\text{with}&&
	\langle q \rangle_t &= \frac{\int q\, S(q,t)\,\mathrm{d}q}{\int S(q,t)\,\mathrm{d}q}
	\;.
\end{align}
\figref{fig_supp_Time_Av_Sq}B shows that $\ell(t)$ initially grows during a transient phase before it fluctuates around a well-defined mean value.
We thus define the pattern length scale $\ell$ as the time average of $\ell(t)$ taken over the stationary regime, excluding the initial transient.
This procedure yields a robust and reproducible length scale that captures the dominant spatial organization of the system despite the absence of a stationary pattern.

\section{Scaling law of length scale in active droplet regime}
\label{app:length_scaling}

\begin{figure}
	\includegraphics[width=\linewidth]{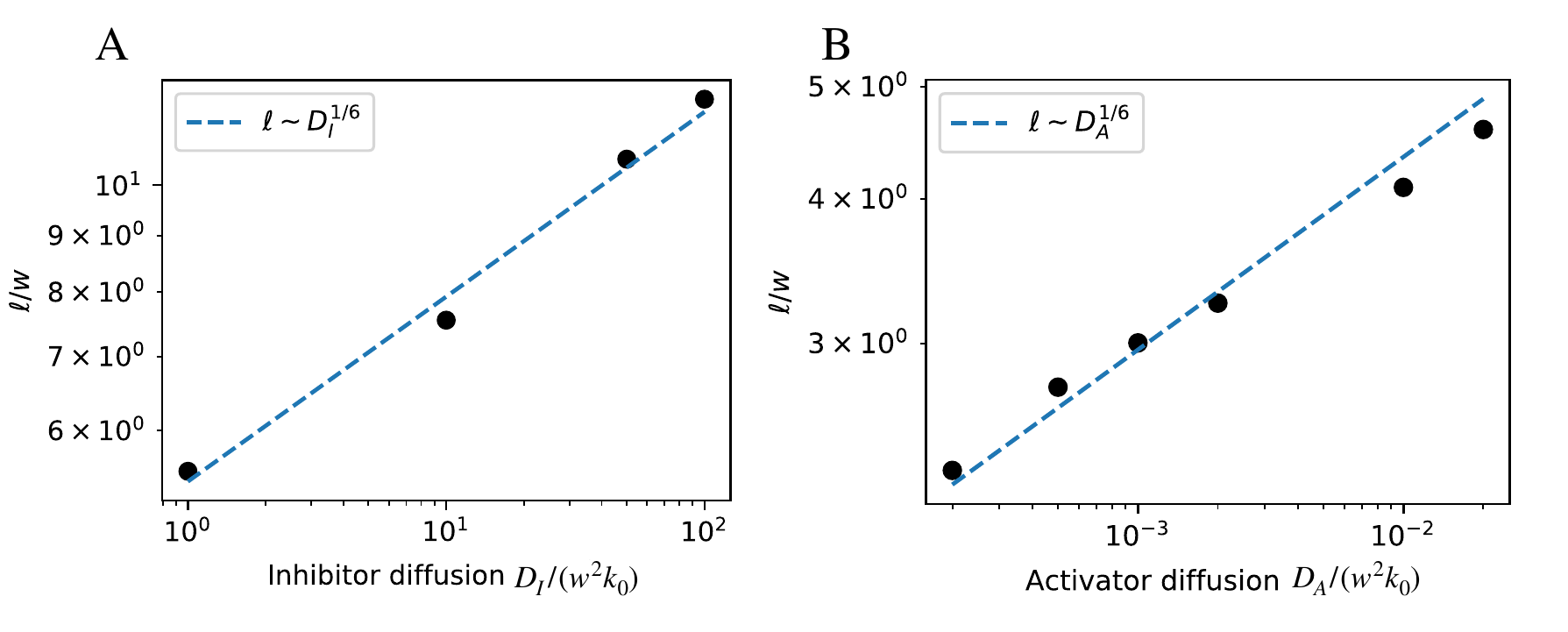}
	\caption{\textbf{Scaling of the pattern length scale in active droplet regime.}  
	Length scale $\ell$ (measured from time-averaged structure factor; see Appendix~\ref{app:length_scale}) as a function of the diffusivity $D_I$ of the inhibitor (panel A, $D_A = 1$) and diffusivity $D_A$ of the activator (panel B, $D_I = 0.1$). Blue dashed lines show the power-law scaling $\ell \sim D_{A/I}^{1/6}$. Model parameters are $\chi = 8$, $h = 5$, $k = 0.1\,k_0$, $\phi_0 = 0.2$, and $k_0 = D_A/w^2$.	}
	\label{fig_supp_Scaling}
\end{figure}

The scaling of the length scale $\ell_\mathrm{AD}$ of  stationary patterns in the active droplet regime is not entirely understood.
To make progress, we use dimensional analysis (i.e., the Buckingham-$\pi$ theorem) to express $\ell_\mathrm{AD}$ as an implicit equation,
\begin{equation}
 F(\ell_\mathrm{AD}/w, D_A/D_I, D_A/(w^2 k), \chi , h) = 0
 \;. 
\end{equation}
Since $\chi$ and $h$ are dimensionless, the length scale reads
\begin{equation} 
	\ell_{\rm AD} = w \cdot  f(l_A/w , l_I/w)
	\;, 
\end{equation}
where $l_i = \sqrt{D_i/k}$ are reaction-diffusion lengths, and $f$ is a non-dimensional scaling function, which may depend on $\chi$ and $h$.
Numerical evidence suggests $\ell\propto D_A^{1/6}$ and $\ell\propto D_I^{1/6}$ (\figref{fig_supp_Scaling}), which implies the power-law $f(x, y) \propto (xy)^{1/3}$ that we propose in section~\ref{sec:AD} of the main text.



\section{Linear stability analysis} 
\label{app:stability_analysis}

\begin{figure*}
\includegraphics[width=\linewidth]{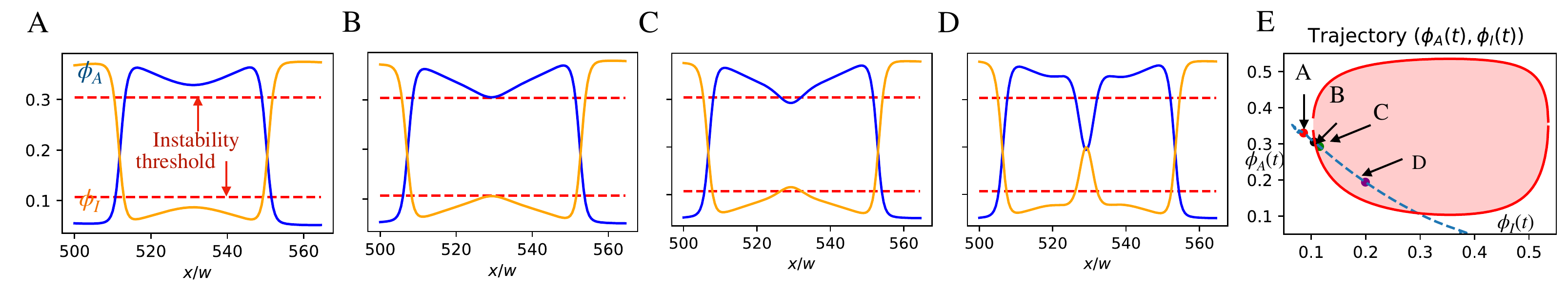}  
	\caption{\textbf{Midpoint trajectory analysis for full model.}
	(A--D) Volume fraction profiles at four times $t$. 
	Red dashed lines mark the instability boundary that is crossed in panel B.
	(E) Phase diagram of system without reactions ($k=0$) as a function of composition ($\phi_A, \phi_I$) obtained from \Eqref{eqn:instability_matrix}.
	Homogeneous states are unstable inside the red region.
	Blue line shows the trajectory at droplet center at $x = 530\,w$ corresponding to panels A--D.
	(A--E) Model parameter are $\chi = 6$, $k = 10^{-3}$, $D_I/D_A = 10$, $h = 5$, $\phi_0 = 0.2$.}
	\label{fig:instability_full_model}
\end{figure*}

We here analyze the linear stability of the dynamical equations \eqref{eqn:pde} for the activator and inhibitor volume fractions $\phi_A(\vect{r}, t)$ and $\phi_I(\vect{r}, t)$ around their homogeneous steady state $\phi_A = \phi_I = \phi_0$. 
To study stability, we linearize the equations around this uniform state and consider small perturbations 
in Fourier space, characterized by the wave vector $\vect{q}$. 
The temporal evolution of these perturbations is governed by the eigenvalues of the Jacobian matrix $J(\vect{q})$, 
which determine whether the homogeneous state is stable or unstable, 
\begin{equation}
\label{eqn:instability_matrix}
J(\vect{q}) =
-\begin{pmatrix}
 \vect{q}^2 D_A  \psi_\mathrm{d} - \frac{(h - 2)k}{2}
 & \vect{q}^2 D_A\psi_\mathrm{od} + \frac{h k}{2} \\ 
 \vect{q}^2 D_I \psi_\mathrm{od} - \frac{h k}{2}
 & \vect{q}^2 D_I \psi_\mathrm{d} + \frac{(h + 2)k}{2}
\end{pmatrix},
\end{equation}
where $\psi_\mathrm{d} = 1 + \psi + w^2 \vect{q}^2$ and $\psi_\mathrm{od}=\psi + \phi_0 \chi$.
The eigenvalues of the Jacobian matrix $J(\vect{q})$ are given by
\begin{equation}
\sigma_{\pm}(\vect{q})
  = \frac{\operatorname{tr} J(\vect{q})}{2}
    \pm \frac{1}{2} 
      \sqrt{\big(\operatorname{tr} J(\vect{q})\big)^{2}
      - 4 \det J(\vect{q})}
	\;,
\label{eq:sigma_pm}
\end{equation}
and an instability occurs when the real part of one of these eigenvalues becomes positive. 
Since the trace is  negative,
\begin{equation}
\operatorname{tr} J(\vect{q}) = -\big[\vect{q}^2(D_A + D_I)(1 + \psi ) + 2k \big] < 0, 
\end{equation}
the eigenvalue $\sigma_{-}(\vect{q})$ always has a negative real part.
Consequently, stability is governed by $\sigma_{+}(\vect{q})$ and thus the sign of $\operatorname{det} J(\vect{q})$.  Without reactions ($k=0$), the system behaves as a passive mixture and $\sigma_{\pm}(\vect{q})$ reduce to purely diffusive eigenmodes, so the condition $\operatorname{det} J(\vect{q}) = 0$ defines the phase boundary separating stable and unstable homogeneous states.
To obtain the stability regions shown in \figref{fig:isolated_droplet}E and \figref{fig:instability_full_model}E, we solve the condition $\operatorname{det} J(\vect{q}) = 0$ to leading order in $\vect q$, where it becomes independent of $|\vect q|$.
We then determine the instability threshold in composition space by solving for $(\phi_I, \phi_A)$ at fixed values of the interaction strength $\chi$.

\section{Midpoint trajectory analysis for full dynamics}
\label{app:instability_full_model}

Section~\ref{sec_DR} showed that tracing the mid point of an isolated droplet in phase space predicts the instability that leads to fission.
\figref{fig:instability_full_model} shows the same analyses for the fully developed case of the dynamic pattern, providing evidence that the qualitative features of the analysis for the single droplet model are reproduced in the full model.

\section{Dynamical patterns can be preceded by long transients}
\label{app:transients}

\begin{figure}
	\includegraphics[width=\linewidth]{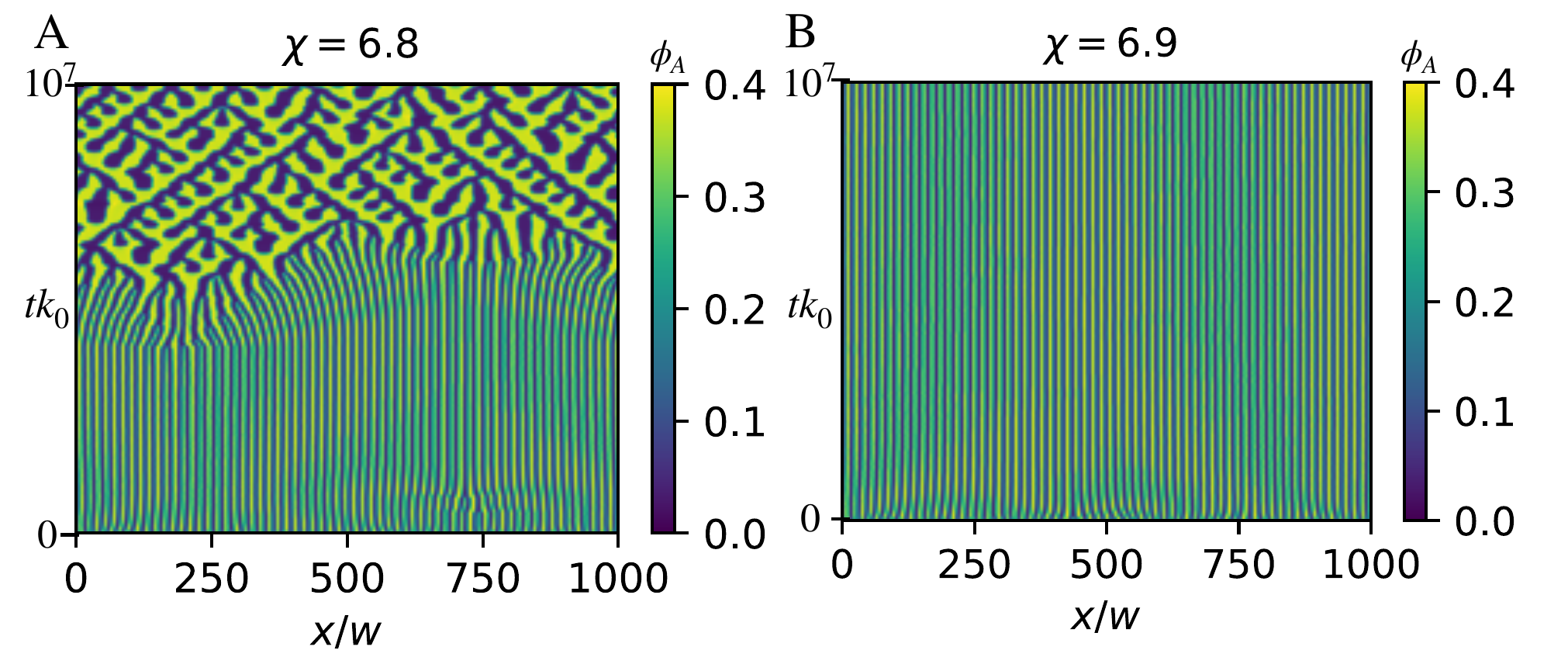}
	\caption{\textbf{Dynamical pattern regime can exhibit long transients.} 
	Volume fraction of activator $\phi_A$ as a function of the one spatial coordinate~$x$ and time $t$ for $\chi = 6.8$ (panel A) and $\chi = 6.9$ (panel B).	Model parameters are $h = 5$, $D_I/D_A = 10$, $k = 0.01\,k_0$, $\phi_0 = 0.2$, and $k_0 = D_A/w^2$. 
	}
	\label{fig:transients}
\end{figure}

\figref{fig:transients} displays additional spacetime plots for the dynamical regime (panel~A, $\chi = 6.8$) and the active droplet regime (panel~B, $\chi = 6.9$).
In particular, the data for $\chi = 6.8$ reveals that the onset of the growth-fission cycle characteristic for the dynamical regime can be preceded by a long transient. 
However, we found early time signatures in all such transient datasets which in \figref{fig:transients}A are visible around $x/w = 250$ and $x/w = 750$. Namely, the larger spacing between the temporal concentration lines that are notably absent in data sets deemed stable (\figref{fig:transients}B).
We therefore consider this an early time signature to the spatio-temporal dynamics and used this fact for classification in \figref{fig:kymographs}(D-H) in case of long transients.

\bibliography{references}

\end{document}